\documentclass[ip,twocolumn]{jpsj3}

\usepackage{txfonts}
\usepackage{bm}
\usepackage{graphicx}
\usepackage{color}

\title{Dynamics of Deformable Active Particles under External Flow Field}

\author{Mitsusuke Tarama}
\inst{Fukui Institute for Fundamental Chemistry, Kyoto University, Kyoto, 606-8103, Japan} 

\abst{
In most practical situations, active particles are affected by their environment, for example, by a chemical concentration gradient, light intensity, gravity, or confinement. 
In particular, the effect of an external flow field is important for particles swimming in a solvent fluid. 
For deformable active particles such as self-propelled liquid droplets and active vesicles, as well as microorganisms such as euglenas and neutrophils, a general description has been developed by focusing on shape deformation. 
In this review, we present our recent studies concerning the dynamics of a single active deformable particle under an external flow field. 
First, a set of model equations of active deformable particles including the effect of a general external flow is introduced. 
Then, the dynamics under two specific flow profiles is discussed: a linear shear flow, as the simplest example, and a swirl flow. 
In the latter case, the scattering dynamics of the active deformable particles by the swirl flow is also considered. 
}%

\begin{document}

\maketitle

\section{Introduction} \label{sec:Introduction}

In the last decade, the dynamics of active particles has attracted much attention from various viewpoints of nonequilibrium science~\cite{Ramaswamy2010The,Cates2012Diffusive,Romanczuk2012Active,Marchetti2013Hydrodynamics}. 
This is a broad field concerning spontaneous dynamical motion accompanied by symmetry breaking, examples of which are found in both biological and artificial systems~\cite{Vicsek2012Collective,Ebbens2010In,Lauga2009The}. 
Some active particles are rigid with a prescribed fixed shape such as self-propelled colloids~\cite{Paxton2004Catalytic,Jiang2010Active,Kapral2013Perspective,Bechinger2016Active} and microswimmers such as bacteria~\cite{Aranson2013Active} and \textit{Chlamydomonas}~\cite{Ishikawa2009Suspension}. 
Deformable active particles also exist, the shape of which changes in time. 
In particular, shape deformation is of great importance for biological organisms such as eukaryotic cells~\cite{Keren2008Mechanism,Li2008Persistent,Maeda2008Ordered,Bosgraaf2009The,Mogilner2009Shape}. 
Deformable active particles are also realized by using liquid droplets and vesicles that undergo chemical reactions on the interface~\cite{Nagai2005Mode,Kitahata2011Spontaneous,Toyota2009Self-propelled,Hanczyc2011Metabolism,Miura2010Autonomous,Ban2013ph-dependent,Giardini2003Compression,Boukellal2004Soft}. 

Depending on the environment, active particles are also classified into two different groups. 
One is swimming-type active particles, which self-propel in a fluid or on the interface between liquid and gas. 
Such swimmers include active colloidal particles and microorganisms in a solvent and camphor solids on the interface of an aqueous phase~\cite{Nakata1997Self-rotation}. 
The other is crawling active particles, which migrate on solid or soft substrates, such as self-propelled liquid droplets~\cite{Sumino2005Self-running,John2005Self-propelled} and eukaryotic cells~\cite{Fournier2010Force,Tanimoto2014A}. 
In the latter case, the adhesion to the substrate plays an important role~\cite{Schwarz2013Physics}.

In order to elucidate the dynamics of these particles from a theoretical viewpoint, one needs to conduct studies of nonlinear dynamics and nonequilibrium statistical physics. 
A number of elaborate models have been studied for each specific system, both for swimming particles~\cite{Golestanian2005Propulsion,Barthes-Biesel2006From,Molina2014Diffusion,Tjhung2012Spontaneous} and for crawling ones~\cite{Carlsson2011Mechanisms,Shao2012Coupling,Recho2013Contraction-Driven,Dreher2014Spiral,Ziebert2016Computaional}, by taking into account the details of internal mechanisms. 
However, since the example of active particles includes both biological and synthetic systems, a basic theoretical description of active particles is also needed. 
For the case of deformable active particles, we have developed a general description~\cite{Ohta2009Deformable,Hiraiwa2010Dynamics,Hiraiwa2011Dynamics,Tarama2012Spinning,Tarama2013DynamicsPTEP,Tarama2013Oscillatory,Tarama2016Reciprocating,Ohta2016Simple}. 
By focusing on shape deformation, we derived nonlinear time-evolution equations from symmetry considerations. 

In contrast to active motion, the classical passive motion of a particle is a mechanical reaction to external forces. 
While the external forces acting on a particle do not vanish when integrated over the whole particle, the force relevant to active motion vanishes, which is known as the force-free and torque-free property of active particles. 
Still, force-free and torque-free active particles undergo a variety of dynamical motion even without external forces~\cite{Purcell1977Life}. 
Here, symmetry breaking plays an important role so that active particles achieve spontaneous motion~\cite{Najafi2004Simple,Gunther2008A}. 

Then, a naturally arising question is what dynamics appear if the active particles move under external stimuli. 
In fact, in most realistic situations, there are various external effects caused by the environment. 
One is chemotaxis~\cite{Bagorda2008Eukaryotic,Friedrich2007Chemotaxis,Hiraiwa2013Theoretical}, where the particles sense a chemical concentration gradient. 
This is particularly relevant to biological cells~\cite{Bagorda2008Eukaryotic}, which migrate either towards the higher chemical concentration or away from it. 
Phototaxis is another important external factor~\cite{Garcia2013Light,Lozano2016Phototaxis} especially for some biological cells~\cite{Jekely2009Evolution}. 
External stimuli can also be mechanical such as gravity~\cite{Tarama2011Dynamics} and confinement~\cite{Berke2008Hydrodynamic,Tailleur2009Sedimentation,Mino2011Enhanced,Elgeti2013Wall,Takagi2014Hydrodynamic,van_Teeffelen2009Clockwise-directional}. 

An important particular external stimulus for swimming active particles is a solvent flow field. 
The simplest external flow profile is a linear shear flow, which has been studied theoretically for rigid circular active particles~\cite{tenHagen2011Brownian} and for deformable active particles~\cite{Tarama2013DynamicsJCP}, as well as experimentally by using bacteria~\cite{Rusconi2014Bacterial}. 
Another flow profile that is more practical is a Poiseuille flow through tubes~\cite{Tao2010Swimming,Zottl2012Nonlinear,Kessler1985Hydrodynamic,Mathijssen2016Upstream}. 
A realistic situation is swirl flows, which occur naturally, including in turbulent systems. 
Interestingly, such a flow is also induced by active particles themselves~\cite{Sokolov2007Concentration,Dunkel2013Fluid}. 
The dynamics of active particles in a swirl flow was investigated theoretically including particle shape deformation~\cite{Tarama2014Deformable}, which was followed by a study of an artificial model swimmer~\cite{Kuchler2016Getting}. It has also been studied experimentally using bacterial suspensions~\cite{Sokolov2016Rapid}.

In this paper, we review our recent studies of active deformable particles in an external flow field. 
The organization of this review is as follows. 
In the next section (Sect.~\ref{sec:Model}), we introduce a set of model equations that describes the dynamics of active deformable particles in a solvent flow field. 
Sections.~\ref{sec:Linear Shear Flow} and \ref{sec:Swirl Flow} are devoted to the dynamics in a linear shear flow~\cite{Tarama2013DynamicsJCP} and in a swirl flow~\cite{Tarama2014Deformable}, respectively. 
In addition to the steady-state solutions, the collision dynamics of active particles with the swirl flow are presented in Sect.~\ref{sec:Swirl Flow}.

\section{Model} \label{sec:Model}

Here, we introduce the equations of motion that describe the dynamics of an active deformable particle under the influence of an external flow field. 
In order to make the analysis general, a set of coupled nonlinear dynamical equations is derived from symmetry considerations. 
These equations are proposed for the general case of three spatial dimensions and an unspecified flow profile. 
Afterwards, we confine ourselves to two dimensions and consider a linear shear flow and a swirl flow. 

First, we consider the flow field without the presence of the active particles. 
For simplicity, we assume that the flow profile is externally imposed and thus prescribed. 
We denote the external flow velocity by $\bm{u}$, which varies as a function of space. 
The spatial dependence is characterized by the elongational part $\mathsf{A}$ and the rotational part $\mathsf{W}$, which are defined by~\cite{Pleiner2004Nonlinear}
\begin{align}
&A_{ij} = ( \partial_i u_j +\partial_j u_i ) /2,
 \label{eq:Aij}\\
&W_{ij} = ( \partial_i u_j -\partial_j u_i ) /2,
 \label{eq:Wij}
\end{align}
where the indices $i$ and $j$ label the Cartesian coordinates. 

An active particle moves spontaneously in the external flow field. 
That is, on the one hand, it is passively advected by the external flow and, on the other hand, it can actively self-propel with respect to the surrounding fluid. 
The active velocity measured with respect to the surrounding fluid flow is denoted by $\bm{v}$. 
Here, we consider a small particle so that the Reynolds number is sufficiently small. 
Then, the time evolution of the center-of-mass position of the particle $\bm{x}$ is given by
\begin{equation}
\frac{d x_i}{dt} = u_i + v_i.
 \label{eq:dx_i/dt}
\end{equation}

In the same manner, the rotation of the particle is divided into two parts. 
One is the passive rotation due to the external flow field, which is denoted by $\mathsf{W}$, 
and the other is the active rotation that the particle exhibits spontaneously, which is denoted by $\mathsf{\Omega}$. 
The total rotation of the particle is represented by
\begin{equation}
\mathsf{W} +\mathsf{\Omega}.
 \label{eq:total_rotation}
\end{equation}
From Eq.~\eqref{eq:Wij}, $\mathsf{W}$ is an antisymmetric tensor, and so is $\mathsf{\Omega}$. 
The antisymmetric tensor of the active rotation is related to the angular velocity vector $\bm{\omega}$ as
\begin{equation}
\Omega_{ij} = \epsilon_{ijk} \omega_k,
 \label{eq:Omega_omega}
\end{equation}
where $\epsilon_{ijk}$ denotes the $(i,j,k)$ component of the Levi-Civita tensor.
Summation over repeated indices is implied, as throughout the remainder of this review. 

Now, we introduce the description of the shape deformation of the particle. 
For simplicity, we first consider two spatial dimensions and then explain the three-dimensional case. 
The shape of the particle is determined by the position of the interface, which is represented by the local radius with respect to the particle center of mass:
\begin{equation}
R(\tilde{\theta},t) = R_0 +\delta R(\tilde{\theta},t),
 \label{eq:R}
\end{equation}
where the angle $\tilde{\theta}$ measures the direction around the center of mass. 
$R_0$ in Eq.~\eqref{eq:R} stands for the equilibrium shape without deformation, for which we assume a circular shape. 
Thus, $R_0$ is a positive constant. 
The deformation is then described by the deviation from the equilibrium circular shape, $\delta R(\tilde{\theta},t)$, which may depend on time.
Here, we assume that the shape deformation is not very large so that the local radius is a single-valued function with respect to the angle $\tilde{\theta}$. 
Generally, the deformation is expanded in a Fourier series as
\begin{equation}
\delta R (\tilde{\theta},t) = \sum_{m=2}^{\infty} \big( z_m(t) e^{im\tilde{\theta}} + z_{-m}(t) e^{-im\tilde{\theta}} \big).
 \label{eq:deltaR}
\end{equation}
Here, the zeroth mode is excluded by assuming that the original circular shape is sufficiently stable.
The first Fourier mode represents the translation of the center of mass, which is therefore included in the center-of-mass velocity $\bm{v}$. 
The lowest-mode deformation is thus given by the second Fourier mode $z_{\pm2}$, which represents an elliptical deformation. 
In two dimensions, we can define the second-rank traceless symmetric tensor as
\begin{equation}
\mathsf{S} = 
\left(
\begin{array}{cc}
S_{11} & S_{12} \\
S_{21} & S_{22}
\end{array}
\right)
= \left(
\begin{array}{cc}
s \cos 2\theta & s \sin 2\theta \\
s \sin 2\theta & -s \cos 2\theta
\end{array}
\right),
 \label{eq:S}
\end{equation}
where we define $z_{\pm2} = (s/2) \exp(\mp 2i \theta)$. 
Note that the symmetric tensor can also be defined for each of the higher-order deformation modes~\cite{Tarama2013Oscillatory,Fel1995Tetrahedral}. 
The symmetric-tensor description of the deformation is general in the sense that the same form is applicable for both two and three dimensions. 
In the case of three spatial dimensions, the deviation $\delta R$ must be expanded into spherical harmonics $Y_{\ell m}$ with coefficients $c_{\ell m}$, to which the symmetric deformation tensor is related likewise~\cite{Hiraiwa2011Dynamics}. 
The lowest mode of the deformation is given by $\ell=2$, representing an ellipsoidal deformation. 
Hereafter, we take into account only the lowest-mode deformation $\mathsf{S}$. 

In total, we have introduced three central dynamical variables to characterize the state of a deformable active particle: 
the active propulsion velocity $\bm{v}$, the active rotation $\mathsf{\Omega}$, and the elliptical deformation $\mathsf{S}$. 
For the sake of generality, the time evolution equations for these variables are derived on the basis of symmetry arguments. 
We consider the following coupled nonlinear equations~\cite{Tarama2013DynamicsJCP}: 
\begin{align}
&\frac{d v_i}{d t} = \alpha v_i -(v_k v_k) v_i -a_1 S_{ik} v_k -a_2 ( W_{ik} +\Omega_{ik} ) v_k, 
 \label{eq:dv_i/dt}\\
&\frac{d \Omega_{ij}}{d t} = \zeta \Omega_{ij} -\mu_0 ( \Omega_{k\ell} \Omega_{k\ell} ) \Omega_{ij} \notag\\
&\hspace{1.2em} +\mu_1 ( \Omega_{ik} S_{kj} - \Omega_{jk} S_{ki} ) +\mu_2 S_{ik} \Omega_{k\ell} S_{\ell j},
 \label{eq:dOmega_ij/dt}\\
&\frac{d S_{ij}}{d t} = -\kappa S_{ij} +b_1 \Big[ v_i v_j -\frac{1}{d} ( v_k v_k ) \delta_{ij} \Big] \notag\\
&\hspace{1.2em} -b_2 [ (W_{ik} +\Omega_{ik}) S_{kj} + (W_{jk} +\Omega_{jk}) S_{ki} ] \notag\\
&\hspace{1.2em} +b_3 \Big[ \Omega_{ik} S_{k\ell} \Omega_{\ell j} -\frac{1}{d} ( \Omega_{mk} S_{k\ell} \Omega_{\ell m} ) \delta_{ij} \Big] +b_4 ( \Omega_{k\ell} \Omega_{k\ell} ) S_{ij} \notag\\
&\hspace{1.2em} +\nu_1 \Big( A_{ij} -\frac{1}{d} A_{kk} \delta_{ij} \Big) +\nu_2 \Big[ A_{ik} S_{kj} +A_{jk} S_{ki} -\frac{2}{d} ( A_{k\ell} S_{\ell k} ) \delta_{ij} \Big].
 \label{eq:dS_ij/dt}
\end{align}
Here, $\delta_{ij}$ denotes the Kronecker delta and $d$ is the dimension of the space. 
All the coefficients are phenomenological parameters. 
The meaning of each term in Eqs.~\eqref{eq:dv_i/dt}--\eqref{eq:dS_ij/dt} is explained below. 
In principle, more terms and higher-order couplings can be included, but the current model already covers the main physical aspects that we intend to describe. 
Note that we consider the effect of the external flow on the particle dynamics but the inverse effect is not included here. 
Such an effect is important for, for example, hydrodynamic interactions between particles~\cite{Molina2014Diffusion}. 
In this paper, we devote ourselves to the single-particle dynamics. 

We start with the first two terms on the right-hand side of Eq.~\eqref{eq:dv_i/dt}. 
They can be rewritten as
\begin{equation}
-\frac{\partial \mathcal{F}}{\partial v_i} ~\textrm{with}~ \mathcal{F} = -\frac{\alpha}{2} (v_k v_k) +\frac{1}{4} (v_k v_k)^2, 
 \label{eq:F}
\end{equation}
where $\mathcal{F}$ is a Lyapunov functional controlling the spontaneous self-propulsion. 
By increasing $\alpha$, the particle exhibits a supercritical pitchfork bifurcation from $\bm{v}=0$ to $\bm{v}\neq0$ at $\alpha=0$, corresponding to the onset of self-propulsion. 
Such a structure is also considered in continuum descriptions of flocks of active particles~\cite{Toner1995Long-Range,Toner1998Flocks,Toner2005Hydrodynamics}. 
The drift bifurcation formula, Eq.~\eqref{eq:F}, together with the time derivative term on the left-hand side of Eq.~\eqref{eq:dv_i/dt}, was derived for the isolated domain solution of reaction-diffusion equations in both two~\cite{Ohta2009Deformation} and three dimensions~\cite{Shitara2011Deformable}. 
It was also derived from the Stokes equation for a droplet catalyzing a chemical reaction on its inside, which changes the interfacial tension~\cite{Yabunaka2012Self-propelled,Yoshinaga2014Spontaneous}. 
The Marangoni flow arising from the nonuniform distribution of the interfacial tension causes the droplet to move. 
The equation for the migration velocity is derived in the limit of an infinitesimally thin interface. 
The time derivative term of the center-of-mass velocity appears as a consequence of the time delay, that is, the relaxation of the concentration is much slower than that of the fluid velocity. 

The same bifurcation structure is taken into account for the active rotation of the particle $\mathsf{\Omega}$ in the first line of Eq.~\eqref{eq:dOmega_ij/dt}. 
Indeed, the spontaneous spinning motion of active particles is found in experiments~\cite{Takabatake2011Spontaneous,Ebata2015Swimming,Tierno2008Controlled,Wang2009Dynamic}. 
The parameter $\zeta$ characterizes the active rotation strength; the particle does not exhibit a spontaneous rotation if $\zeta\le0$; otherwise, it spins with angular velocity $\sqrt{\zeta/2\mu_0}$. 
Here, $\mu_0$ is a positive constant. 

In contrast, a spontaneous deformation of the particle is not considered here. 
The linear damping term with coefficient $\kappa>0$ in Eq.~\eqref{eq:dS_ij/dt} describes the relaxation of the deformation back to the spherical (circular) shape.  The deformation is then caused through the coupling terms to $\bm{v}$ and $\bm{\Omega}$ and by stretching due to the external flow. 

The elongational part of the externally imposed flow field $\mathsf{A}$ deforms the particle shape through the terms with coefficients $\nu_1$ and $\nu_2$. 
These terms are identical to those in a previous study on the dynamics of a nonactive droplet in a fluid flow, which also considered elliptical shape deformation~\cite{Stark2003Poisson-bracket,Maffettone1998Equation}. 
Note that the term with coefficient $\nu_2$ vanishes for a two-dimensional incompressible flow.

The deformation is also induced by the active velocity through the second term on the right-hand side of Eq.~\eqref{eq:dS_ij/dt}. 
The coefficient $b_1$ determines the magnitude of the effect and also the tendency of the elongation direction. 
When $b_1$ is positive (negative), the particle tends to elongate parallel (perpendicular) to the self-propulsion direction. 
The third term on the right-hand side of Eq.~\eqref{eq:dv_i/dt} is the counter term of the $b_1$ term if $a_1 b_1 <0$. 
That is, both terms are derived from the functional derivative of $b_1 v_k S_{k\ell} v_{\ell}$~\cite{Tarama2014Individual}. 
However, here we take the same sign for $a_1$ and $b_1$ considering the nonvariational case.
Then, because of the $a_1$ term, the self-propulsion speed changes and its direction reorients depending on the particle deformation. 
These terms are the leading-order coupling terms between the velocity $\bm{v}$ and the deformation $\mathsf{S}$. 

The effect of these coupling terms between the active velocity and second-order deformation was studied previously by using Eqs.~\eqref{eq:dv_i/dt} and \eqref{eq:dS_ij/dt} without the active rotation ($\mathsf{\Omega}=\mathsf{0}$)~\cite{Ohta2009Deformable}. 
After undergoing the bifurcation from the motionless state $\bm{v}=\bm{0}$ without any deformation $\mathsf{S}=\mathsf{0}$ to the self-propulsion regime $\bm{v}\neq\bm{0}$ at $\alpha=0$, the particle migrates spontaneously in a straight trajectory with its shape elliptically deformed $\mathsf{S}\neq\mathsf{0}$.  
The straight solution is stable as long as $0<\alpha<\alpha^*$, while it becomes unstable and the particle starts to migrate in a circular trajectory for $\alpha>\alpha^*$. 
Here the bifurcation threshold from the straight motion to the circular motion is given by~\cite{Ohta2009Deformable}
\begin{equation}
\alpha^* = \frac{\kappa^2}{a_1 b_1} +\frac{\kappa}{2}. 
 \label{eq:alpha*}
\end{equation}

The active rotation $\mathsf{\Omega}$ also influences the deformation $\mathsf{S}$ through the coupling terms with coefficients $b_3$ and $b_4$ on the right-hand side of Eq.~\eqref{eq:dS_ij/dt}. 
For $b_3>0$ and $b_4>0$, the spontaneous rotation of the particle enhances the degree of the deformation, while it reduces it for $b_3<0$ and $b_4<0$. 
In two dimensions, these terms are equivalent for $b_3=2b_4$. 
In a three-dimensional space, the term with coefficient $b_3$ includes an additional effect that rotates the deformed particle, in contrast to the term with $b_4$. 
Note that the $b_3$ term is corrected in Eq.~\eqref{eq:dS_ij/dt}, which was not traceless in three dimensions in the previous formula in Ref.~\citenum{Tarama2013DynamicsJCP}. 
This correction, however, does not change any results discussed in Refs.~\citenum{Tarama2013Dynamics JCP} and \citenum{ Tarama2014Deformable} since we considered only two-dimensional dynamics. 
The third and fourth terms on the right-hand side of Eq.~\eqref{eq:dOmega_ij/dt} have a similar effect on the rotation $\mathsf{\Omega}$ induced by the deformation $\mathsf{S}$. 
Note that the term with coefficient $\mu_1$ vanishes in two dimensions. 

Finally, both the active rotation of the particle $\mathsf{\Omega}$ and the passive rotation due to the external flow field $\mathsf{W}$ reorient the particle active velocity and the particle configuration. 
This effect is included by the term with coefficient $a_2$ in Eq.~\eqref{eq:dv_i/dt} and by the term with coefficient $b_2$ in Eq.~\eqref{eq:dS_ij/dt}. 
Since an active particle can follow a prescribed rule on how to react to an external rotational flow, the numerical value of the coefficients cannot be fixed at this point. 
In principle, the contribution with $a_2>0$ describes so-called Magnus effect, a force acting on the particle in the direction perpendicular to the velocity and angular velocity. 

Generally, the model equations of Eqs.~\eqref{eq:dx_i/dt} and \eqref{eq:dv_i/dt}--\eqref{eq:dS_ij/dt} apply to a three-dimensional setup.
For simplicity, however, we confine ourselves to two spatial dimensions for the remainder of this review.

\section{Linear Shear Flow} \label{sec:Linear Shear Flow}

As the simplest example of a flow profile, a linear steady shear flow is considered in two spatial dimensions. 
The flow velocity is given by
\begin{equation}
\bm{u} = ( \dot{\gamma} y, 0 ),
 \label{eq:shear}
\end{equation}
where $\dot{\gamma}$ is the shear rate. 

For the analytical investigation below, we parametrize the position, the velocity, and the angular velocity by $\bm{x}=(x,y)$,
\begin{align}
\bm{v} = ( v\cos\phi, v\sin\phi ),~
\mathsf{\Omega} = \left(
\begin{array}{cc}
0 & \omega \\
-\omega & 0
\end{array}
\right).
 \label{eq:coordinate}
\end{align}
Then, Eqs.~\eqref{eq:dx_i/dt} and \eqref{eq:dv_i/dt}--\eqref{eq:dS_ij/dt} are rewritten as
\begin{align}
&\frac{d x}{dt} = v\cos\phi +\dot{\gamma} y,~
\frac{d y}{dt} = v\sin\phi,
 \label{eq:shear:dposition/dt}\\
&\frac{dv}{dt} = \alpha v -v^3 -a_1 v s \cos 2(\theta-\phi), 
 \label{eq:shear:dv/dt}\\
&\frac{d\phi}{dt} = -a_1 s \sin 2(\theta-\phi) +a_2 \Big( -\frac{\dot{\gamma}}{2} +\omega \Big),
 \label{eq:shear:dphi/dt}\\
&\frac{d\omega}{dt} = \zeta \omega - 2\mu_0 \omega^3 -\mu_2 s^2 \omega,
 \label{eq:shear:domega/dt}\\
&\frac{ds}{dt} = -\kappa s +\frac{b_1 v^2}{2} \cos 2(\theta-\phi) +\tilde{b} s \omega^2 +\frac{\nu_1 \dot{\gamma}}{2} \sin2\theta,
 \label{eq:shear:ds/dt}\\
&\frac{d\theta}{dt} = -\frac{b_1 v^2}{4 s} \sin 2(\theta-\phi) +b_2 \Big( -\frac{\dot{\gamma}}{2} +\omega \Big) +\frac{\nu_1 \dot{\gamma}}{4 s} \cos 2\theta, 
 \label{eq:shear:dtheta/dt}
\end{align}
where we define $\tilde{b} = b_3 +2 b_4$. 

From these expressions, it is obvious that both $x$ and $y$ do not affect the dynamics of the other variables, and so we can solve Eqs.~\eqref{eq:shear:dv/dt}--\eqref{eq:shear:dtheta/dt} separately from Eq.~\eqref{eq:shear:dposition/dt}. 
The dynamics governed by Eqs.~\eqref{eq:shear:dv/dt}--\eqref{eq:shear:dtheta/dt}, and thus, by Eqs.~\eqref{eq:dv_i/dt}--\eqref{eq:dS_ij/dt} with Eq.~\eqref{eq:shear}, are invariant under the simultaneous transformations $(a_1,b_1)\rightarrow(-a_1,-b_1)$ and $\phi \rightarrow \phi +\pi/2$. 
This invariance ensures that the choice of signs for $a_1$ and $b_1$ does not change the dynamical structure. 
As we have explained in Sect.~\ref{sec:Model}, the choice of signs determines the characteristic direction of the self-propulsion with respect to that of the elliptical deformation; the particle tends to self-propel in the longitudinal (lateral) direction of the elliptic shape if $a_1>0$ and $b_1> 0$ (if $a_1<0$ and $b_1<0$). 
The dynamics of the particle position, Eq.~\eqref{eq:shear:dposition/dt}, is given by a superposition of the dynamics governed by Eqs.~\eqref{eq:shear:dv/dt}--\eqref{eq:shear:dtheta/dt} and the simple advection due to the external flow $\bm{u}$. 
Therefore, although the particle trajectory is shifted slightly, the dynamical structures such as the transition from one dynamical solution to another are not affected by the choice of the signs of $a_1$ and $b_1$. 

In the following, we first consider the limited case of a circular shape without deformation to see the role of the particle activity and also to clarify the connection to a model of a rigid active particle. 
Then, we describe the dynamics of a self-propelled deformable particle but without active rotation.

\subsection{Rigid active particle} \label{sec:Rigid active particle}

Here, we consider the special case of a rigid active particle of circular shape. 
We neglect the deformation, and thus we set $s=0$ in Eqs.~\eqref{eq:shear:dposition/dt}--\eqref{eq:shear:domega/dt}, and drop Eqs.~\eqref{eq:shear:ds/dt} and \eqref{eq:shear:dtheta/dt} from our model equations. 

We now assume that the magnitude of the velocity $v$ and that of the relative rotation $\omega$ relax quickly so that they are given by the steady-state solutions of Eqs.~\eqref{eq:shear:dv/dt} and \eqref{eq:shear:domega/dt}, i.e., $v=v_0$ and $\omega=\pm\omega_0$, where 
\begin{align}
&v_0 = \sqrt{\alpha},
 \label{eq:rigid:v}\\
&\omega_0 = \sqrt{\zeta /2\mu_0}.
 \label{eq:rigid:omega}
\end{align}
The positive (negative) sign in front of $\omega_0$ corresponds to the counterclockwise (clockwise) rotation. 

For these solutions, Eqs.~\eqref{eq:shear:dposition/dt} and \eqref{eq:shear:dphi/dt} are solved as
\begin{align}
&x(t) = v_0 ( \tilde{\omega} -\dot{\gamma} / \tilde{\omega}^2 ) [ \sin\phi(t) -\sin\phi_0 ] \notag\\
&\hspace{2.4em} +\dot{\gamma} [ (v_0 / \tilde{\omega} ) \cos\phi_0 +y_0 ] t +x_0,
 \label{eq:rigid:x(t)}\\
&y(t) = - (v_0 /\tilde{\omega}) [ \cos\phi(t) -\cos\phi_0 ] +y_0,
 \label{eq:rigid:y(t)}\\
&\phi(t) = \tilde{\omega} t +\phi_0,
 \label{eq:rigid:phi(t)}  
\end{align}
where $(x_0,y_0)$ and $\phi_0$ are the position of the center of mass and the direction of the velocity vector at $t=0$, respectively. 
This set of solutions represents a cycloidal trajectory as long as $\tilde{\omega}\neq 0$. 
Here, we have defined the effective angular velocity
\begin{equation}
\tilde{\omega} = a_2 [ -(\dot{\gamma}/2) \pm \omega_0 ].
 \label{eq:rigid:omega-tilde}
\end{equation}

In the special case that the effective angular velocity vanishes, $\tilde{\omega}=0$, the passive rotation due to the external flow is balanced by the particle active rotation. 
Then, Eq.~\eqref{eq:shear:dphi/dt} gives $\phi(t) = \phi_0$ and the solutions of Eq.~\eqref{eq:shear:dposition/dt} become
\begin{align}
&x(t) = ( \dot{\gamma} v_0 /2) ( \sin\phi_0 ) t^2 +( v_0 \cos\phi_0 +\dot{\gamma} y_0 ) t +x_0,
 \label{eq:rigid:x(t)_norotation}\\
&y(t) = ( v_0 \sin\phi_0) t + y_0.
 \label{eq:rigid:y(t)_norotation}
\end{align}
The meanings of $(x_0,y_0)$ and $\phi_0$ are the same as above. 
This set of solutions causes the particle to move in a parabolic trajectory instead of a cycloid. 

Similar equations of motion and results were also obtained for
an active Brownian particle
under a linear shear flow~\cite{tenHagen2011Brownian}. 
In this case, the equations of motion are in the overdamped limit. 
The particle possesses a polarity, along which it tends to self-propel. 
Its self-propulsion speed as well as its active rotation, which rotates the polarity, fluctuate around constant values. 
In the limit of no fluctuation, the same trajectories as above are obtained.

\subsection{Active deformable particle} \label{sec:Active deformable particle}

Now, we consider the dynamics of a self-propelled deformable particle in a linear shear flow.
For simplicity, we include only the active velocity and eliminate the active rotation by setting $\zeta<0$. 
\begin{figure*}[tb]
\begin{center}
\includegraphics[width=0.85\textwidth]{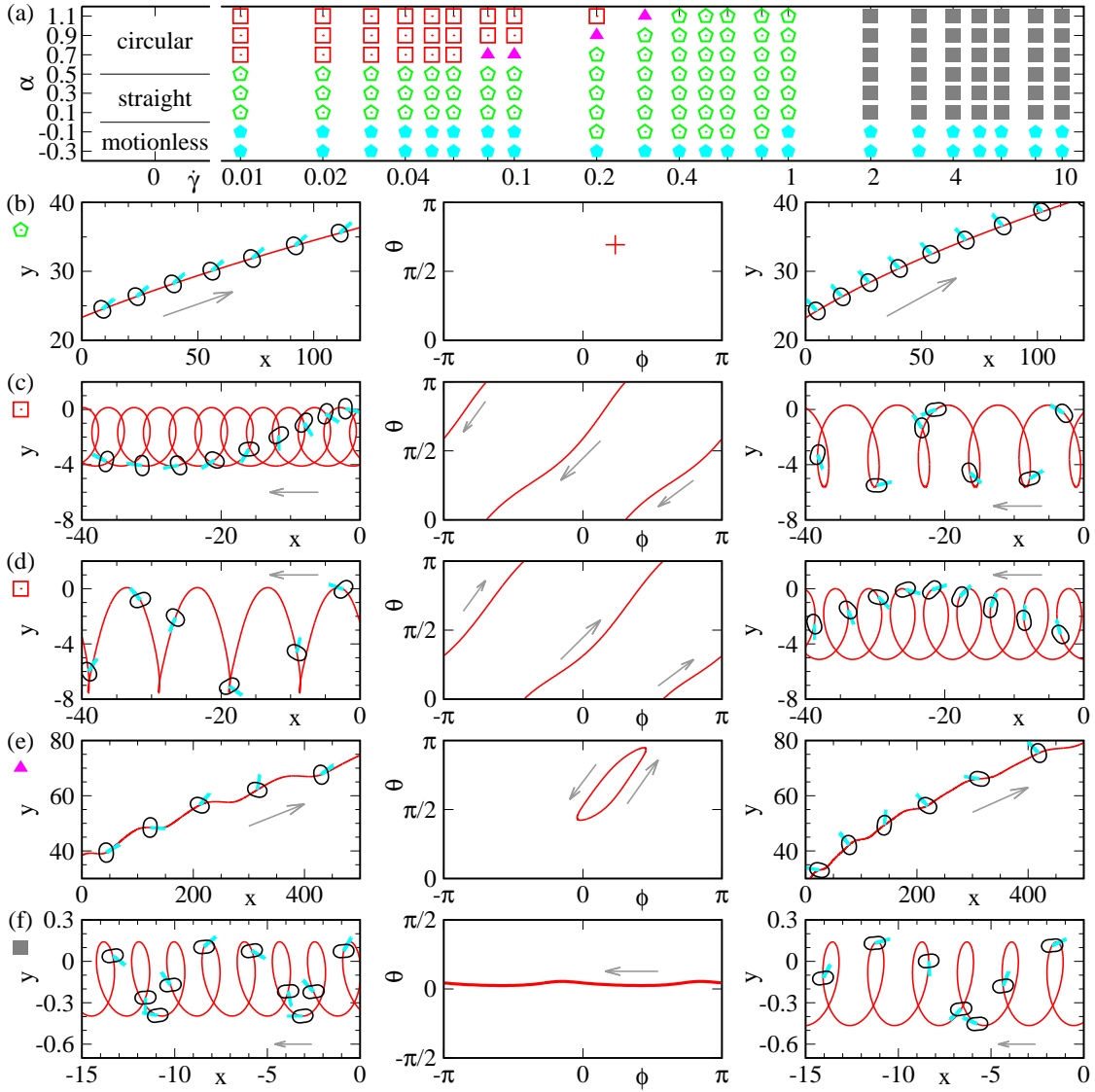}
\caption{(color online)
(a) Dynamical phase diagram and (b)--(f) trajectories in real space  (left and right columns) and attractors in $\theta$-$\phi$ space (middle column) of typical dynamical motion; 
(b) active straight motion for $\alpha=0.5$ and $\dot{\gamma}=0.1$, indicated by the green open pentagons in panel (a); 
(c) and (d) cycloidal I motion with the rotation of the particle deformation in the clockwise and counterclockwise directions, respectively, for $\alpha=0.9$ and $\dot{\gamma}=0.1$, marked by the red open squares in panel (a); 
(e) winding motion for $\alpha=0.7$ and $\dot{\gamma}=0.08$, indicated by the purple filled triangles in panel (a); 
(f) cycloidal motion II without the rotation of the particle deformation for $\alpha=0.1$ and $\dot{\gamma}=2$, marked by the gray filled squares in panel (a).
Arrows in panels (b)--(f) show the direction of motion. 
Some snapshots of the particle silhouette, the size of which is adjusted for illustration, are superimposed onto the real-space trajectories. 
The cyan bars on the particle silhouettes denote the direction of the active velocity $\bm{v}/v$. 
In panels (b)--(f), the results in the left and middle columns are obtained for the perpendicular particles with $a_1=b_1=-1$, whereas those in the right column are for the parallel particles with $a_1=b_1=1$. 
In panel (a), cyan filled pentagons represent the passively advected state without active velocity $v=0$. 
The active spinning $\mathsf{\Omega}$ equals $\mathsf{0}$ for all of these examples of motion~\cite{Tarama2013DynamicsJCP}. 
}
\label{fig:shear_deformable}
\end{center}
\end{figure*}
We numerically integrate Eqs.~\eqref{eq:dx_i/dt} and \eqref{eq:dv_i/dt}--\eqref{eq:dS_ij/dt} with $\zeta=-0.1$. 
In the following simulation results, we numerically confirmed that the active rotation always vanishes ($\mathsf{\Omega}=0$). 
Here, we discuss the perpendicular case with $a_1=b_1=-1$. 
The other parameters are chosen as $a_2=b_2=\mu_2=\nu_1=1$ and $\tilde{b}=1$. 
The results of the numerical simulation are summarized in Fig.~\ref{fig:shear_deformable}. 
Panel~(a) shows the dynamical phase diagram, where the shear rate $\dot{\gamma}$ of the imposed linear shear flow and the magnitude of the particle self-propulsion $\alpha$ are varied for the intermediate damping rate of deformation $\kappa_2=0.5$. 
Panels~(b)--(f) display the trajectories in real space (left and right columns) and the attractors in $\phi$-$\theta$ space for each dynamical solution. 
The figures in the right column are obtained for $a_1=b_1=1$ and the others are for $a_1=b_1=-1$. 

As mentioned in Sect.~\ref{sec:Model}, there are three solutions in the absence of an external flow: the motionless solution for $\alpha<0$, the straight solution for $0<\alpha<\alpha^*$, and the circular solution for $\alpha^*<\alpha$. 
From Eq.~\eqref{eq:alpha*}, $\alpha^*=0.5$ for our present parameters. 
These solutions are indicated by ``motionless'', ``straight'', and ``circular'' in Fig.~\ref{fig:shear_deformable}(a), respectively. 

If the external shear flow is switched on, the particle in the motionless state $\alpha<0$ then exhibits a trivial solution, where it is slightly elongated and simply advected parallel to the flow for the parameters indicated by the cyan filled pentagons in Fig.~\ref{fig:shear_deformable}(a). 
For positive $\alpha$, the particle self-propels on top of the passive advection. 
If $0<\alpha<\alpha^*$, it undergoes active straight motion, as denoted by the green open pentagons in Fig.~\ref{fig:shear_deformable}(a). 
For this solution, the active velocity is finite and time-independent, and thus the trajectory becomes straight if seen from the frame comoving with the flow, as shown in Fig.~\ref{fig:shear_deformable}(b). 
On the other hand, the particle describes a cycloidal trajectory for $\alpha>\alpha^*$, where the shape deformation rotates with its center-of-mass trajectory. 
The cycloidal trajectories with counterclockwise and clockwise rotations are depicted in Figs.~\ref{fig:shear_deformable}(c) and \ref{fig:shear_deformable}(d), respectively. 
We refer to these solutions as cycloidal I motion to distinguish them from the other cycloidal solution that we explain shortly. 
The parameter region where the cycloidal I motion was obtained is plotted by the red open squares in Fig.~\ref{fig:shear_deformable}(a). 
As the shear rate increases, the cycloidal I motion rotating in the opposite direction to the external flow, i.e., in the counterclockwise direction, first becomes unstable, and then that with the clockwise rotation loses its stability and starts to undergo winding motion, as shown by the purple filled triangles in Fig.~\ref{fig:shear_deformable}(a). 
Unlike the cycloidal I motion, the shape deformation of the particle for the winding solution does not rotate but oscillates as displayed in Fig.~\ref{fig:shear_deformable}(e). 
The magnitude of the oscillation decreases with increasing $\dot{\gamma}$ and finally vanishes so that the particle undergoes the active straight motion.  
Note that there is a wide range of $\dot{\gamma}$ where the active straight solution was obtained for $\alpha>0$ in Fig.~\ref{fig:shear_deformable}(a). 
For a much larger shear rate, the particle with $\alpha>0$ exhibits the cycloidal II motion, where its shape is always elongated almost horizontally and does not rotate, as displayed in Fig.~\ref{fig:shear_deformable}(f). 
This solution is found for the shear rates denoted by the gray filled squares in Fig.~\ref{fig:shear_deformable}(a). 

As noted at the beginning of Sect.~\ref{sec:Linear Shear Flow}, Eqs.~\eqref{eq:shear:dv/dt}--\eqref{eq:shear:dtheta/dt} can be solved independently of Eq.~\eqref{eq:shear:dposition/dt} and they are invariant with respect to the simultaneous transformations $(a_1,b_1)\rightarrow(-a_1,-b_1)$ and $\phi\rightarrow\phi+\pi/2$. 
Indeed, the numerical simulation of Eqs.~\eqref{eq:dx_i/dt} and \eqref{eq:dv_i/dt}--\eqref{eq:dS_ij/dt} with $a_1=b_1=1$ results in the same dynamical phase diagram as Fig.~\ref{fig:shear_deformable}(a), which was obtained for $a_1=b_1=-1$. 
However, the real-space trajectories are slightly modified as shown in the plots in the right column of Figs.~\ref{fig:shear_deformable}(b)--\ref{fig:shear_deformable}(f), which should be compared with those in the left column of Figs.~\ref{fig:shear_deformable}(b)--\ref{fig:shear_deformable}(f). 
Here, all the other parameters are kept the same. 

On the whole, for small shear rates, the particle undergoes the motion that is obtained as the superposition of the passive advection due to the flow and the active motion, which it exhibits in the absence of the external flow. 
As the shear rate increases, the effect of the external shear flow increases and the dynamics becomes complicated. 
This feature is the same even if the active rotation exists $\mathsf{\Omega}\neq\mathsf{0}$~\cite{Tarama2013DynamicsJCP}.

\section{Swirl Flow} \label{sec:Swirl Flow}

Next, we consider the dynamics of an active deformable particle under a rotational flow (swirl). 
Swirl flows occur naturally in many situations including turbulence. 
The flow velocity of the swirl that we consider here is of the form
\begin{equation}
\bm{u} = \big( - \sigma y /(x^2 +y^2), \sigma x /(x^2 +y^2) \big),
 \label{eq:swirl_u}
\end{equation}
where $\sigma$ sets the strength of the vortex. 
Since this flow profile possesses rotational symmetry, we measure the particle center-of-mass position by
\begin{equation}
\bm{x} = ( r \cos \eta, r \sin \eta )
 \label{eq:swirl_x}
\end{equation}
with distance $r$ and direction $\eta$ with respect to the center of the vortex flow, whereas the center-of-mass velocity and deformation are parametrized by Eq.~\eqref{eq:coordinate}. 
Note that the vortex flow given by Eq.~\eqref{eq:swirl_u} has a flow potential such that $\bm{u} = - \nabla U$, where $U = \mu \arctan ( x/y )$. 
Consequently, $\nabla \times \bm{u} = \bm{0}$, which implies that there is no local rotational contribution, i.e., $\mathsf{W} = \mathsf{0}$. 
In contrast, the stretching contribution does not vanish and is calculated as
\begin{equation}
\mathsf{A} = \left(
\begin{array}{cc}
\sigma r^{-2} \sin 2\eta & -\sigma r^{-2} \cos 2\eta \\
-\sigma r^{-2} \cos 2\eta & -\sigma r^{-2} \sin 2\eta
\end{array}
\right).
 \label{eq:swirl_A}
\end{equation}
From this, one can see that the equations of the active velocity [Eq.~\eqref{eq:dv_i/dt}] and the shape deformation [Eq.~\eqref{eq:dS_ij/dt}] depend on the position of the particle in the case of the swirl, unlike the case of the linear shear flow in the previous section. 
This implies that the dynamics of the parallel configuration ($a_1,b_1>0$) and the perpendicular configuration ($a_1,b_1<0$) likely differ. 

We first investigate the steady-state solutions of Eqs.~\eqref{eq:dx_i/dt} and \eqref{eq:dv_i/dt}--\eqref{eq:dS_ij/dt} with Eq.~\eqref{eq:swirl_u} by numerically integrating them. 
Afterwards, we consider the scattering dynamics of active deformable particles by the swirl flow. 
The setup of a swirl has a geometrical similarity to that of a collision and scattering experiment and therefore possesses a resemblance to the classical Kepler and Rutherford problem. 
We distinguish the two cases of self-propulsion in the parallel and perpendicular directions with respect to the elongation of the particle shape.

\subsection{Steady-state solutions} \label{sec:steady-state}

First, we discuss the steady-state solutions of an active deformable particle in the swirl flow. 
The model equations are too complicated to solve analytically, and therefore, we numerically integrate Eqs.~\eqref{eq:dx_i/dt} and \eqref{eq:dv_i/dt}--\eqref{eq:dS_ij/dt} with Eq.~\eqref{eq:swirl_u}. 
Since the flow profile [Eq.~\eqref{eq:swirl_u}] possesses rotational symmetry, we only vary the initial distance from the flow center.
Here, we distinguish the particles that tend to align the active velocity parallel to the elongation of the shape deformation and those that tend to self-propel perpendicularly. 
The parallel particles are realized by setting $a_1=b_1=1$, and the perpendicular ones by $a_1=b_1=-1$. 
In both cases, the other parameters are fixed as $\kappa=0.5$, $\nu_1=1$, and $\sigma=1$. 
Note that there are no contributions from the terms with coefficients $a_2$, $b_2$, and $\tilde{b}$ since we omit the active rotation, i.e., $\mathsf{\Omega}=\mathsf{0}$, and the swirl flow [Eq.~\eqref{eq:swirl_u}] does not possess any rotational contribution $\mathsf{W}=\mathsf{0}$.

\begin{figure}[tb]
\begin{center}
\includegraphics[width=0.8\columnwidth]{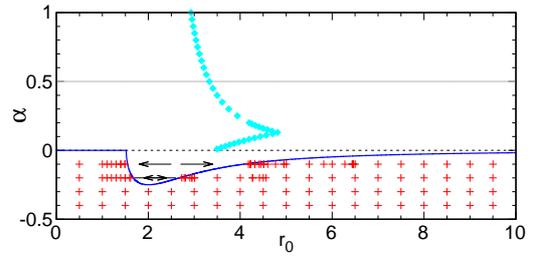}
\caption{(color online)
Radius of stable circular trajectories around the vortex center for the perpendicular configuration ($a_1=b_1=-1$)~\cite{Tarama2014Deformable}. 
}
\label{fig:swirl_passive}
\end{center}
\end{figure}
Before discussing the case of active particles, we consider the motion of a passive particle in the swirl, i.e. $\alpha<0$. 
In this case, the particle is always passively advected by the circular flow, following a circular trajectory around the vortex center. 
The circular trajectory is marginally stable in the radial direction, that is, its radius can take any value depending on the initial conditions. 
In Fig.~\ref{fig:swirl_passive}, the red pluses indicate the radius of the marginally stable passive circular motion, which are obtained starting from different initial distances. 
However, if $\alpha$ becomes close to the bifurcation threshold of self-propulsion located at $\alpha=0$, a region appears where the particles cannot stay and where they are repelled from, as shown by the horizontal gray arrows in Fig.~\ref{fig:swirl_passive}. 
The theoretical analysis~\cite{Tarama2014Deformable} reveals that the radius of the passive rotation $r_0$ should satisfy the stability condition $r_0\le r_{\rm min}$ or 
\begin{equation}
\alpha +2\sigma^2 (r_{\rm min}^{-4} -r_0^{-4} )^{1/2} (\kappa^2 r_0^4 +4 \sigma^2)^{-1/2} < 0
 \label{eq:swirl_radius_threshold}
\end{equation}
for $r>r_{\rm min}$. 
Here, we have defined
\begin{equation}
r_{\rm min} = ( 2 |\sigma| )^{1/2} [ (a_1 \nu_1)^2 -\kappa^2 ]^{-1/4}.
 \label{eq:swirl_r_min}
\end{equation}
In this passive case, there is no difference between the parallel case $a_1,b_1>0$ and the perpendicular case $a_1,b_1<0$. 
However, a difference appears if the particle possesses an active velocity, i.e., $\alpha>0$, as we will see next. 

\begin{figure}[tb]
\begin{center}
\includegraphics[width=\columnwidth]{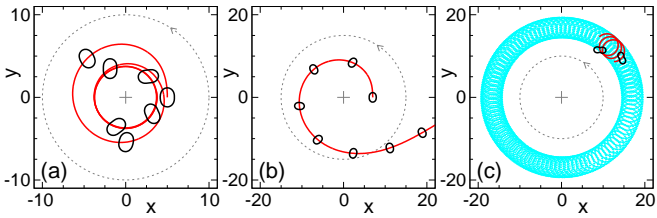}
\caption{(color online)
Trajectories of the particle in real space for $a_1=b_1=-1$~\cite{Tarama2014Deformable}. 
(a) Active circular motion for $\alpha=0.3$, 
(b) escaping motion for $\alpha=0.3$, and
(c) lunar-type motion $\alpha=1$. 
Some snapshots of the particle silhouettes are superposed onto the trajectories. 
In panel (c), the trajectory for longer time intervals is plotted in cyan. 
The rotational flow profile is displayed in gray circular arrows. 
The plus indicates the flow center.
}
\label{fig:swirl_perpendicular}
\end{center}
\end{figure}

First, we discuss the perpendicular case ($a_1=b_1=-1$), where the active velocity tends to align perpendicular to the elongation direction of the shape deformation. 
When $0<\alpha<\alpha^*$, where $\alpha^*=0.5$ for the current parameters, the particle undergoes circular motion around the vortex center, as displayed in Fig.~\ref{fig:swirl_perpendicular}(a). 
We refer to this as active circular motion because, in contrast to the passive circular motion that appears for $\alpha<0$, only one radius of the circular trajectory $r_0$ is selected for each self-propulsion strength $\alpha$, as shown by the cyan diamonds in Fig.~\ref{fig:swirl_passive}. 
Depending on the initial conditions, the swimmer either asymptotically approaches this orbit or it manages to escape from the swirl to an infinite distance, as shown in Fig.~\ref{fig:swirl_perpendicular}(b). 

When $\alpha> \alpha^*$, the situation becomes markedly different. 
Starting sufficiently close to the radius $r_0$, we still observe the steady-state active circular motion as indicated in Fig.~\ref{fig:swirl_passive}. 
However, another type of motion occurs depending on the initial conditions. 
We refer to it as lunar-type motion, the typical trajectory of which is depicted in Fig.~\ref{fig:swirl_perpendicular}(c) for $\alpha=1$.
This trajectory is understood as the circular motion that already occurs in the absence of the swirl for $\alpha>\alpha^*$~\cite{Ohta2009Deformable} superimposed onto the circular convection due to the vortex flow. 
In this case, both rotation directions, the one of the smaller revolution and the one of the larger revolution, are the same as that of the fluid flow. 
The radius of the larger revolution depends on the initial conditions.

\begin{figure}[tb]
\begin{center}
\includegraphics[width=\columnwidth]{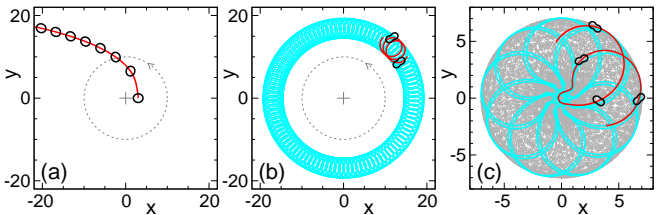}
\caption{(color online)
Trajectories of the particle in real space for $a_1=b_1=1$~\cite{Tarama2014Deformable}. 
(a) Escaping motion for $\alpha=0.1$, 
(b) lunar-type motion for $\alpha=1$, and
(c) multicircular motion $\alpha=1$.
Trajectories for a longer time are plotted in cyan. 
In panel (c), a trajectory with a much longer time is shown in gray.
See the caption of Fig.~\ref{fig:swirl_perpendicular} for a detailed explanation. 
}
\label{fig:swirl_parallel}
\end{center}
\end{figure}

On the other hand, the active circular motion is not found in the parallel case ($a_1=b_1=1$), where the particle tends to self-propel in the elongation direction of the shape deformation. 
Instead, all particles escape far from the vortex center when $0<\alpha<\alpha^*=0.5$, as shown in Fig.~\ref{fig:swirl_parallel}(a). 
In contrast, for $\alpha>\alpha^*$, a particle again undergoes lunar-type motion as displayed in Fig.~\ref{fig:swirl_parallel}(b). 
However, the smaller revolution and the fluid flow have opposite directions of rotation, whereas the rotation directions of the larger revolution and the fluid flow are identical. 
Again, the radius of the larger revolution depends on the initial conditions.

When $\alpha>0.7$, the situation becomes more complex in the parallel case. 
Depending on the initial conditions, multicircular motion can emerge as illustrated in Fig.~\ref{fig:swirl_parallel}(c), where the lighter gray, cyan, and red lines show trajectories of different time intervals. 
To obtain the multicircular motion, the swimmer was initially placed relatively close to the vortex center.

In summary, in a swirl flow, the difference between the parallel ($a_1=b_1=1$) and perpendicular ($a_1=b_1=-1$) configurations has a strong effect on the steady-state solutions, unlike the case of the linear shear flow shown in Sect.~\ref{sec:Linear Shear Flow}. 
The active deformable particles with the perpendicular configuration either escape from the swirl flow or are captured depending on the initial conditions. 
In the latter case, they exhibit the active circular motion around the vortex center for $0<\alpha<\alpha^*$ and the lunar-type motion for $\alpha>\alpha^*$. 
Active deformable particles with the parallel configuration always escape from the swirl for $0<\alpha<\alpha^*$, while for $\alpha>\alpha^*$, they undergo the lunar-type motion or, if they are initially placed very close to the vortex center, the multicircular motion.

\subsection{Scattering dynamics} \label{sec:scattering}

Now, we study the collision dynamics of the active deformable particles with the swirl flow. 
This is performed in analogy to a classical scattering experiment. 
As will be explained shortly, the particles are either scattered or captured by the swirl. 
If the particles are scattered and manage to escape from the vortex, we measure the scattering angle of the event. 
For this purpose, we determine the scattering angle $\eta_{\rm scat}$ between the initial velocity orientation and the final velocity orientation when the particle has reached a certain distance $r_{\rm scat}$ from the vortex center. 
Owning to the swirl geometry, the event of passing the vortex center on one side differs from that of passing it on the other side. 
Therefore, in the following numerical simulations, we measure the scattering angles $\eta_{\rm scat}$ by integrating the changes in the particle velocity during the course of scattering. 

To make the setup meaningful in the sense of a scattering experiment, we set the propulsion strength to values $0<\alpha<\alpha^*$. 
For these values, the particle undergoes straight motion in the absence of the flow field~\cite{Ohta2009Deformable}. 
We provide this solution as an initial condition and place the particle at a comparatively large distance $r_{\rm init} = 1.5 \times 10^4$, with its active velocity heading towards the vortex center. 
If the particle was not affected by the flow field of the swirl, it would propel exactly in the direction of its initial velocity orientation. 
The distance $d_{\rm imp}$ by which it would then miss the vortex center is called the impact parameter. 
A swimmer of $d_{\rm imp}=0$ would hit the center of the vortex if it were not affected by the swirl flow. 
We define $d_{\rm imp} >0$ when the particle velocity is initially oriented towards the side of the oppositely directed fluid flow. 
In contrast, we set $d_{\rm imp} < 0$ when the particle initially propels towards the side of the identically directed fluid flow. 
See Figs.~\ref{fig:swirl_scattering_perpendicular}(b) and \ref{fig:swirl_scattering_parallel}(b) for an illustration of the definition of the sign of $d_{\rm imp}$. 

After numerically integrating Eqs.~\eqref{eq:dx_i/dt}, \eqref{eq:dv_i/dt}, and \eqref{eq:dS_ij/dt} with Eq.~\eqref{eq:swirl_u}, we measure the scattering angle at the distance $r_{\rm scat} = 10^4$ if a scattering event occurs. 
We varied the values of the propulsion strength $\alpha$ and the impact parameter $d_{\rm imp}$, while the other parameters were chosen as before. 
Our results are summarized in Figs.~\ref{fig:swirl_scattering_perpendicular} and \ref{fig:swirl_scattering_parallel}, which display the scattering angles as functions of the impact parameter in panel (a) and the real-space trajectories for $\alpha=0.3$ on a large scale in panel (b) and on a small scale in the vicinity around the swirl center in panels (c) and (d). 
Again, we distinguish between the perpendicular configuration ($a_1 = b_1 = -1$) and the parallel configuration ($a_1 = b_1 = 1$).

\subsubsection{Perpendicular configurations} \label{sec:scattering_perpendicular}

\begin{figure*}[tb]
\begin{center}
\includegraphics[width=\textwidth]{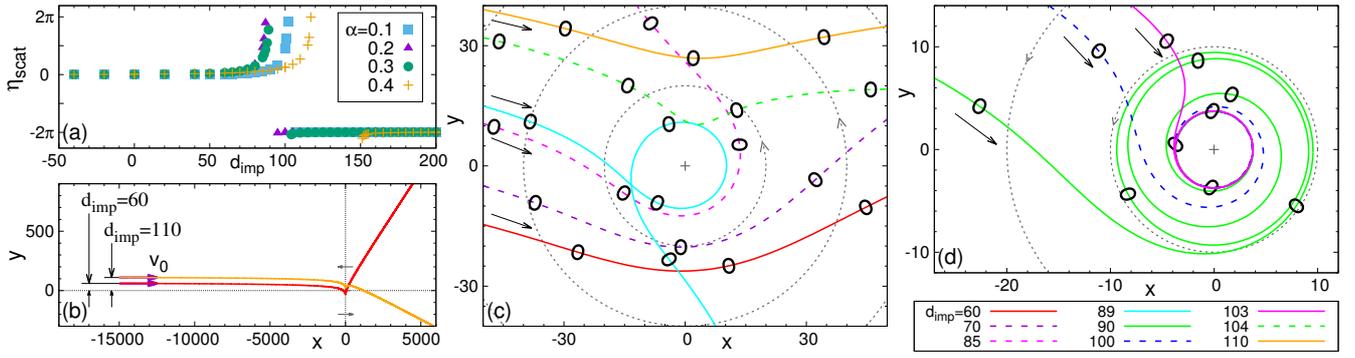}
\caption{(color online)
Scattering dynamics for the perpendicular case with $a_1=b_1=-1$~\cite{Tarama2014Deformable}. 
(a) Scattering angle $\eta_{\rm scat}$ as a function of the impact parameter $d_{\rm imp}$ for various propulsion strengths $\alpha$. 
Real-space trajectories (b) on the large scale including the initial part of the scattering process and (c,d) around the flow center for different impact parameters. 
Some snapshots of the particle silhouettes are superposed onto the trajectories. 
In panel (b), the scales of the $x$- and $y$-axes are chosen differently for illustration. 
The rotational flow profile is displayed by gray arrows in panel (b) and by gray circular arrows in panels (c) and (d). 
}
\label{fig:swirl_scattering_perpendicular}
\end{center}
\end{figure*}

First, we show the collision of the active deformable particles with the perpendicular configuration with the swirl. 
Generally, the vortex causes the particle trajectory to deviate from the original straight one, as in Fig.~\ref{fig:swirl_scattering_perpendicular}(b). 

For negative impact parameters $d_{\rm imp}$, the particle is only weakly deformed and stays far from the vortex center so that it finally leaves the vortex with basically the same velocity orientation as the incident velocity. 
Thus, the scattering angle $\eta_{\rm scat}$ becomes almost zero, as plotted in Fig.~\ref{fig:swirl_scattering_perpendicular}(a). 
This is true even for slightly positive impact parameters, as illustrated in Fig.~\ref{fig:swirl_scattering_perpendicular}(a) and
 Figs.~\ref{fig:swirl_scattering_perpendicular}(b) and \ref{fig:swirl_scattering_perpendicular}(c) for $d_{\rm imp} = 60$. 
With increasing impact parameter, the scattering angle increases, as shown in Fig.~\ref{fig:swirl_scattering_perpendicular}(a), where the particle approaches even closer to the vortex center and its trajectory becomes significantly bent as displayed in Fig.~\ref{fig:swirl_scattering_perpendicular}(c) for $d_{\rm imp} = 70$ and 85. 
The particle circles around the vortex center before it escapes from the swirl as depicted in Fig.~\ref{fig:swirl_scattering_perpendicular}(c) for $d_{\rm imp} = 89$. 
This strong effect of the swirl on the particle dynamics is indicated by the scattering angles $\eta_{\rm scat}>\pi$ in Fig.~\ref{fig:swirl_scattering_perpendicular}(a). 

With further increasing $d_{\rm imp}$, the scattering angle seems to diverge in Fig.~\ref{fig:swirl_scattering_perpendicular}(a). 
Indeed, at even higher impact parameters, the particle is eventually caught by the swirl and cannot escape from it, as shown in Fig.~\ref{fig:swirl_scattering_perpendicular}(d) for $d_{\rm imp} = 90$, 100, and 103. 
Interestingly, in all these cases, the trajectories end in the same circle around the vortex center. 
This attractive trajectory corresponds to the active circular motion discussed in Sect.~\ref{sec:steady-state}.

Finally, when the impact parameter is very large, the particle no longer moves sufficiently close to the swirl center to be effectively captured. 
Instead, it is scattered again; however, it passes the vortex center on the other side, where it propels against the flow velocity
as shown in Fig.~\ref{fig:swirl_scattering_perpendicular}(c) for $d_{\rm imp} = 104$ and 110. 
This scattering event where the particle propels in the opposite direction to the flow velocity is indicated by the scattering angle $\eta_{\rm scat}$ approximately $-2\pi$ in Fig.~\ref{fig:swirl_scattering_perpendicular}(a).

The stability of the active circular motion and the robustness of the capturing event were studied numerically by introducing a stochastic noise term to the equation of the active velocity [Eq.~\eqref{eq:dv_i/dt}]~\cite{Tarama2014Deformable}. 
The latter was also investigated by changing the initial distance, which reveals that the active deformable particles are captured in qualitatively the same manner independent of the initial distance. 
Although the active circular motion is stable for large fluctuation intensities, the capturing event turns out to be much more fragile. 
This is because the trajectory can be shifted considerably owning to the fluctuation on the path towards the swirl center so that the particle does not even reach the close vicinity of the vortex center to be captured.

\subsubsection{Parallel configurations} \label{sec:scattering_parallel}

Next, we consider the collision of the particles with the parallel configuration $a_1=b_1=1$. 
In this case, no permanent capturing by the swirl was observed, in contrast to the case of the perpendicular configuration. 
\begin{figure*}[tb]
\begin{center}
\includegraphics[width=\textwidth]{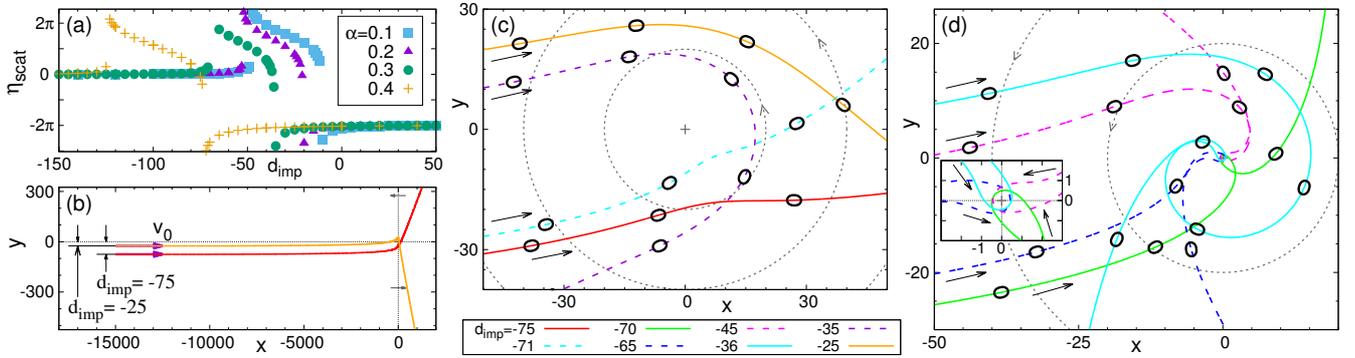}
\caption{(color online)
Scattering dynamics for the parallel case ($a_1=b_1=1$)~\cite{Tarama2014Deformable}. 
(a) Scattering angle $\eta_{\rm scat}$ as a function of the impact parameter $d_{\rm imp}$ for different $\alpha$ values. 
Real-space trajectories (b) on the large scale and (c,d) around the flow center for different $d_{\rm imp}$ values. 
In the inset of panel (d), trajectories closer to the flow center are displayed. 
See the caption of Fig.~\ref{fig:swirl_scattering_perpendicular} for a further explanation. 
}
\label{fig:swirl_scattering_parallel}
\end{center}
\end{figure*}

While they are heading towards the vortex, the situation for active particles with the parallel configuration is much different from those with the perpendicular configuration; the trajectory becomes curved in the opposite direction [compare Fig.~\ref{fig:swirl_scattering_parallel}(b) with Fig.~\ref{fig:swirl_scattering_perpendicular}(b)]. 
Therefore, significant scattering now takes place for negative impact parameters $d_{\rm imp}$, as demonstrated in Figs.~\ref{fig:swirl_scattering_parallel}(b)--\ref{fig:swirl_scattering_parallel}(d).

For negative impact parameters $d_{\rm imp}$ of large magnitude, the particle trajectory is only slightly affected by the swirl. 
The propulsion direction suffers a slight change, i.e., $\eta_{\rm scat}$ is almost zero, as displayed in Fig.~\ref{fig:swirl_scattering_parallel}(c) for $d_{\rm imp}=-75$. 
With increasing impact parameter, the particle appraoches closer to the vortex center and the scattering angle $\eta_{\rm scat}$ increases; see the trajectory for $d_{\rm imp}=-71$ in Fig.~\ref{fig:swirl_scattering_parallel}(c).

Interestingly, as the impact parameter is increased, the scattering angle jumps discontinuously as in Fig.~\ref{fig:swirl_scattering_parallel}(a). 
The trajectory for $d_{\rm imp}=-70$ in Fig.~\ref{fig:swirl_scattering_parallel}(d) shows the drastic event that occurs in this case, which explains the jump in the scattering angle. 
The particle becomes close to the vortex center and is temporarily caught by the swirl, describing a loop around the vortex center before it escapes. 
As indicated by the trajectories for the other $d_{\rm imp}$ in Fig.~\ref{fig:swirl_scattering_parallel}(d), the same behavior is obtained upon further increasing the impact parameter, although the scattering angle decreases continuously. 
As highlighted by the inset of Fig.~\ref{fig:swirl_scattering_parallel}(d), in all these cases, the particle performs a loop around the center in the same direction as the fluid flow. 

Finally, another discontinuous jump of the scattering angle occurs in Fig.~\ref{fig:swirl_scattering_parallel}(a) at even higher impact parameters. 
After the jump, the particle no longer performs a narrow loop around the vortex center. 
Instead, its trajectory features a simple bend around the swirl, as depicted in Fig.~\ref{fig:swirl_scattering_parallel}(c) for $d_{\rm imp}=-35$ and -25. 
The complete trajectory of the scattering event for $d_{\rm imp}=-25$ is depicted in Fig.~\ref{fig:swirl_scattering_parallel}(b). 
Again, the scattering events where the particle propels against the flow velocity are indicated by the shift of the scattering angle $\eta_{\rm scat}$ by $-2\pi$ in Fig.~\ref{fig:swirl_scattering_parallel}(a).

To sum up, the collision dynamics of active deformable particles with the swirl is divided into two: weak collisions, where the particles are slightly affected but are finally scattered by the swirl, and strong collisions, where the particles reach the close vicinity of the vortex center. 
The weak collisions occur for small and large impact parameters, whereas the strong collisions are observed for intermediate impact parameters. 
In the case of the perpendicular configuration ($a_1=b_1=-1$), the particles with small and large impact parameters are simply advected in addition to the self-propulsion before being scattered from the vortex center as in Figs.~\ref{fig:swirl_scattering_perpendicular}(b) and \ref{fig:swirl_scattering_perpendicular}(c). 
For positive intermediate impact parameters, strong collisions are obtained, where the particles are captured by the swirl as in Fig.~\ref{fig:swirl_scattering_perpendicular}(d). 
In contrast, for the parallel configuration ($a_1=b_1=1$), the particles self-propel even against the flow velocity as shown in Figs.~\ref{fig:swirl_scattering_parallel}(b) and \ref{fig:swirl_scattering_parallel}(c) so that a strong collision is obtained for negative impact parameters. 
In this case, the particles describes a small loop around the vortex center in the same direction as the flow velocity and are finally scattered far away by the swirl as depicted in Fig.~\ref{fig:swirl_scattering_parallel}(d).

\section{Summary} \label{sec:Summary}

We have reviewed the dynamics of active deformable particles under an external flow field. 
By focusing on the shape deformation, we have introduced a general model, which includes the effect of an external flow, based on symmetry considerations. 
Our model does not depend on any details of the specific system such as a mechanical origin of the activity. 
Since the variables are described by tensors, the obtained formulae are applicable to both two and three spatial dimensions. 

We have investigated the dynamics in two dimensions under two different flow profiles, a linear shear flow as the simplest case and a swirl flow. 
In both cases, the time-evolution equations are solved numerically. 
In addition, the dynamics of the collision between an active particle and the swirl flow was investigated numerically, revealing a capturing event and a complicated scattering trajectory depending on the impact parameters.

An important flow that has not yet been studied for the deformable active particles is a Poiseuille flow. 
This type of flow profile often appears in tubes; therefore, the interaction with the boundary wall plays a major role~\cite{Zottl2013Periodic,Mathijssen2016Upstream}. 
Additionally, when the density of the suspension is high, the interaction between the active deformable particles needs to be taken into account. 
Moreover, hydrodynamic interactions can be included either by solving the full fluid dynamic equations~\cite{Zottl2014Hydrodynamics,Oyama2016Purely} or by employing the Green's function method~\cite{Lopez2014Dynamics}. 

Our predictions can be tested in experiments on active liquid droplets floating on the interface of an aqueous phase~\cite{Nagai2005Mode,Kitahata2011Spontaneous}. 
In the case of a three-dimensional space, liquid droplets~\cite{Toyota2009Self-propelled,Hanczyc2011Metabolism} and vesicles~\cite{Miura2010Autonomous,Ban2013ph-dependent} that self-propel in solution were experimentally realized. 
The euglenoid movement of \textit{Eutreptiella gymnastica}~\cite{Farutin2013Amoeboid} and the swimming motion of dictyostelium cells and neutrophils~\cite{Barry2010Dictyostelium,Bae2010On} are biological examples, where shape deformation plays an important role in the particle swimming motion in solution. 
Furthermore, an anomalous change in viscosity is known for suspensions of bacteria~\cite{Hatwalne2004Rheology,Sokolov2009Reduction,Marcos2012Bacterial,Gachelin2013Non-Newtonian,Lopez2015Turning,Figueroa-Morales2015Living,Clement2016Bacterial} which hardly change their shape. 
It is therefore interesting to examine the impact of the shape deformation of active particles on the rheological properties.


\begin{acknowledgments}

The work reviewed in this article was carried out together with T.~Ohta, H.~L\"owen, A.~M.~Menzel, B.~ten~Hagen, and R.~Wittkowski.

\end{acknowledgments}


\end{document}